\documentclass[conference]{IEEEtran}
\IEEEoverridecommandlockouts
\usepackage{cite}
\usepackage{amsmath,amssymb,amsfonts}
\usepackage{algorithmic}
\usepackage{graphicx}
\usepackage{textcomp}
\usepackage{xcolor}
\usepackage{url}
\def\BibTeX{{\rm B\kern-.05em{\sc i\kern-.025em b}\kern-.08em
    T\kern-.1667em\lower.7ex\hbox{E}\kern-.125emX}}
\usepackage[normalem]{ulem}

\newcommand{\Abasis}{\mathcal{A}}

\newcommand{\band}{\&}
\newcommand{\bor}{|}
\newcommand{\bxor}{\oplus}
\newcommand{\bsl}{\ll}
\newcommand{\bsr}{\gg}
\newcommand{\bcl}{\leftarrow}
\newcommand{\bcr}{\rightarrow}
\newcommand{\fqm}{\mathbb{F}_{q^m}}
\newcommand{\fq}{\mathbb{F}_{q}}
\newcommand{\vect}[1]{\boldsymbol{{#1}}}
\newcommand{\mat}[1]{\boldsymbol{{#1}}} 
\begin{document}

\title{On Software Implementation of Gabidulin Decoders}

\author{\IEEEauthorblockN{Johannes Kunz, Julian Renner, Georg Maringer, Thomas Schamberger, Antonia Wachter-Zeh}
\IEEEauthorblockA{\textit{Technical University of Munich (TUM)},
Munich, Germany \\
\{johannes.f.kunz, julian.renner, georg.maringer, t.schamberger, antonia.wachter-zeh\}@tum.de}
\thanks{The work of J. Renner and A. Wachter-Zeh was supported by the European Research Council (ERC) under the European Union’s Horizon 2020 research and innovation programme (grant agreement no. 801434).

G. Maringer's work was supported by the German Research Foundation (Deutsche Forschungsgemeinschaft, DFG) under Grant No. WA3907/4-1.}
}

\maketitle
\begin{abstract}
This work compares the performance of software implementations of different Gabidulin decoders. The parameter sets used within the comparison stem from their applications in recently proposed cryptographic schemes. The complexity analysis of the decoders is recalled, counting the occurrence of each operation within the respective decoders. It is shown that knowing the number of operations may be misleading when comparing different algorithms as the run-time of the implementation depends on the instruction set of the device on which the algorithm is executed.
\end{abstract}


\section{Introduction}
In this work we are considering different decoding approaches for Gabidulin codes. These codes are of special interest since they belong to the class of maximum rank distance (MRD) codes. This work considers the specific parameter sets used within the cryptosystem RQC \cite{rqc}. This cryptographic scheme is a Round 2 candidate in the NIST-PQC competition, which standardizes post-quantum secure cryptographic algorithms. Apart from the desired security level, the performance of the algorithms plays an important role in the standardization process.
The Gabidulin decoder is a major part of the decryption process of the RQC algorithm, hence, it is of particular relevance.

In this work we review the complexity analysis of different decoding algorithms, which is based on counting the number of operations. By implementing the decoders in C, we show that counting the number of theoretically required operations does not give the full picture as the mapping of operations onto the instruction set of the microprocessor by the compiler may significantly change the performance evaluation. In fact operations which are seemingly negligible regarding their complexity within the decoder can play an important role for the performance of the decoding algorithms.

The decoding algorithms discussed within this work are the  Welch-Berlekamp Algorithm (WBA) that is currently implemented in RQC\cite{augot} and the Transform Domain Decoder (TDD) \cite{silva} by D. Silva and F. R. Kschischang.
The two decoders use different basis representations for elements in the finite extension field. While the cryptosystem RQC  performs operations in a polynomial basis, we implement a low-complexity normal basis for the TDD. Both implementations are written in C and are compiled using gcc for the x86-64 instruction set.

\section{Preliminaries}\label{s-preliminaries}
\subsection{Notation}
For the sake of clarity we define the following notation:
Lower-case and upper-case symbols in bold font, e.g. $\vect{a}$ and $\mat{A}$ denote vectors and matrices, respectively. The subscripts $\vect{a}_i$ or $\mat{A}_{i,j}$ are used to indicate the $i$th element of the vector or matrix element at row $i$ and column $j$, accordingly.
We denote the finite field of characteristic $q$ by $\fq$ and its extension of degree $m$ by $\fqm$. Powers of the characteristic (q-powers) are abbreviated by $[i]:= q^i$.
The symbol $\fq[x]$ is the polynomial ring over $\fq$ and $\langle\Pi\rangle$ is the ideal generated by the polynomial $\Pi$. The operator $\text{Tr}(x) = \sum_{k=0}^{m-1}x^{[k]}$ is the trace and $\delta$ denotes the Kronecker delta.

For describing operations in the finite field of characteristic 2, we introduce the binary operators $\band$, $\bor$, and $\bxor$ on vectors, which denote the element-wise AND, OR, and XOR operation, respectively. The left-shift operator $\bsl$ and the right-shift operator $\bsr$ perform non-cyclic shifts filling with zeros while removing entries on the other side of the vector.
In case of $64$bit vectors, these five operators correspond to the bit-wise operators in C. Conversely, the operators $\bcr$ and $\bcl$ denote cyclic shift to left and the right, respectively.

\subsection{Operations in Finite Extension Fields}
The extension field $\fqm$ is a vector space over $\mathbb{F}_q$ that is spanned by the basis $\Abasis = \{\alpha_0, \dots, \alpha_{m-1}\}$. Any element $a \in \fqm$ can be represented as a vector $\vect{a} \in \fq^{1\times m}$. Let $\vect{\alpha} = [\alpha_0, \dots, \alpha_{m-1}]$ be a row vector containing the basis elements. It holds that $a =  \vect{a} \vect{\alpha}^T$.

In practice, operations on finite extension field elements are performed on the vector representation, e.g. on $\vect{a}$. Depending on the choice of $\Abasis$ these operations differ. In the following, we will discuss the properties of operations in polynomial and normal bases and assert their complexity.

\subsubsection{Polynomial Bases}\label{s-preliminaries-poly}
Let $\Pi\in\fq[x]$ be an irreducible polynomial of degree $m$ and $\fqm := \fq[x] /\langle\Pi\rangle$. A polynomial basis is of the form $\{\alpha^{0},\hdots,\alpha^{m-1}\}$, where $\alpha$ is a root of the irreducible polynomial $\Pi\in\fq[x]$ \cite{mullen07}. An element $a\in\fqm$ can be interpreted as a polynomial of degree smaller than $m$, where $\vect{a}$ is the coefficient vector of the polynomial representation.

Let $a$ and $b$ denote two elements of $\fqm$.
Those elements can be added by element-wise addition of their vector representation $\vect{a}$ and $\vect{b}$. This requires $m$ additions in $\fq$.

There exist different algorithms for the multiplication in $\fqm$. We choose an algorithm that is efficient in case $\Pi$ is sparse.
The multiplication is divided into two steps. First, we compute the unreduced product of the two polynomials $a$ and $b$, which has at most degree $2m-2$. This takes $m^2$ multiplications and $(m-1)^2$ additions in $\fq$ \cite{erdem}.
Second, we reduce the product modulo the irreducible polynomial $\Pi$. In case $\Pi$ is a trinomial, i.e. of the form $x^m + x^k + 1, 0<k<m$, the reduction requires $2m-2$ coefficient additions in $\fq$ \cite{erdem}.

In case $q=2$, computing the square of a field element can be done more efficient compared to multiplying the element with itself. Squaring is achieved by inserting zeros in between every two bits of the input coefficient vector. The resulting zero-interleaved coefficient vector is then reduced by the irreducible polynomial \cite{mahboob}. In theory, zero-interleaving does not afford any computation, hence, the total cost of $2m-2$ additions in $\mathbb{F}_2$ is due to the reduction.

As we are dealing with polynomials, inversion can be performed using the extended euclidean algorithm (EEA). The complexity of the EEA depends on the input polynomial, see \cite{hankerson}, Section 2.3.6. Thus, we will simply denote the average required number of additions in $\fq$ as $C_{\text{inv}}$.  
\subsubsection{Normal Bases}\label{s-preliminaries-normal}
A normal basis is defined by 
$\Abasis=\{\alpha^{[0]},\hdots,\alpha^{[m-1]}\}$. Every normal basis has its unique dual basis $\Bar{\Abasis} = \{\Bar{\alpha}_0 \dots \Bar{\alpha}_{m-1}\}$ satisfying $\text{Tr}(\alpha_i \Bar{\alpha}_j) = \delta_{ij}$.

Computing the q-power of $a\in\fqm $ corresponds to a cyclic shift of $\vect{a}$, i.e. $a^{[i]} = \vect{a}^{\bcl i} \vect{\alpha}^T$. As no arithmetic operations are performed, the cost of taking the q-power is assumed to be negligible \cite{silva}.

The multiplication of $\vect{a}$ and $\vect{b}$ is given by
\begin{equation} \label{f-norm-mul}
   \vect{c} = \sum_{i=0}^{m-1}\vect{b}_i (\mat{M}\vect{a}^{\bcl i})^{\bcr i}.
\end{equation}
The matrix $\mat{M} \in \fq^{m\times m}$ is called multiplication table and is sparse in the ideal case. The number of non-zero entries $C_M$ of $\mat{M}$ is the complexity of the basis and is lower bound by $C_M \geq 2m-1$. In case $C_M = 2m-1$, $\Abasis$ is called optimal \cite{gao, silva}. Depending on $m$ there may not exist an optimal basis. In this case one can use a low-complexity normal basis, see \cite{gao, ash}. Consequentially, the complexity of multiplication in normal bases varies with $C_M$ and requires $m^2$ multiplications and $m(C_M - 1)$ additions in $\fq$. 

In one of the decoding algorithms, we frequently perform the multiplication of a field element $a$ with a q-power of the normal element $\alpha$, i.e. $c = a\alpha^{[i]}$. As the representation vector of $\alpha^{[i]}$ contains a single non-zero element at the position $i$, Equation \eqref{f-norm-mul} simplifies to $\vect{c} = (\mat{M}\vect{a}^{\bcl i})^{\bcr i}$ requiring only $C_M - m$ additions in $\fq$.

Finally, multiplicative inverses can be obtained in multiple ways. There are methods based on Fermat's little theorem $a^{-1} = a^{q^m - 2}$, see \cite{itoh}, and algorithms based on the extended euclidean algorithm \cite{sunar}. Our algorithm is inspired by \cite{itoh} and decomposes the power $a^{q^m - 2}$ similarly to \cite{mahmoud}. It is presented in detail for $m=127$ in Section \ref{s-impl-normal-fqm} requiring nine multiplications in $\fqm$.

\subsection{Linearized Polynomials}\label{s-preliminaries-q-poly}
A linearized polynomial (q-polynomial) is defined as \cite{ore}
\begin{equation*}
A(x) = \sum\limits_{i=0}^{n}\vect{a}_ix^{[i]},
\end{equation*}
where $n=\text{deg}_q(A)$ denotes the q-degree of $A$ and $\vect{a}_i\in\fqm$.

Let $\vect{a}$ denote the coefficient vector of $A, \text{deg}_q(A)<m$ zero padded to length $m$, i.e., $\vect{a} = [\vect{a}_0, \dots, \vect{a}_n, \vect{a}_{n+1}, \dots, \vect{a}_{m-1}]$, $\vect{a}_{n+1}, \dots, \vect{a}_{m-1} = 0$. We define a cyclic indexing for the vector elements, i.e., $\vect{a}_{i\mod m} =  \vect{a}_{i}$.

We define the q-transform $\tilde{A}$ of a linearized polynomial with respect to the normal element $\alpha$ by the transformation of the coefficient vector elements, i.e., $\vect{\tilde{a}}_i = \sum_{j=0}^{m-1}\vect{a}_j\alpha^{[i+j]}, i=0\dots m-1$. The q-transform is a linear bijection between the time domain and the transform domain. It can be reversed by performing the q-transform with respect to the dual element $\bar{\alpha}$ of $\alpha$. In case the normal element is self-dual, the inverse q-transform simply is the forward q-transform.

\subsection{Gabidulin Codes}\label{s-preliminaries-gabidulin}
A Gabidulin code $\mathcal{C}$ is a $(n, k)$ block code with a generator matrix $\mat{G} = [\vect{g}_j^{[i]}] \in \fqm^{k \times n}$, $0\leq i <  k, 0 < j  \leq n$, where the generating elements $\vect{g}_j \in \fqm$ have to be linearly independent over $\fq$. Gabidulin codes satisfy the Singleton bound with rank distance $d = n-k+1$. Thus, they can correct up to $\tau_{\text{max}} = \lfloor\frac{n-k}{2}\rfloor$ errors \cite{gabidulin}. The parity check matrix has the structure $\mat{H} = [\vect{h}_j^{[i]}] \in \fqm^{n-k \times n}$, $0\leq i <  n-k, 0\leq j < n$, where $\vect{h}_i$ are linearly independent over $\fq$ \cite{richter}.

\section{Implementation of Finite Extension Field Operations in C}\label{s-impl-fqm}
In this section, we describe the software implementation of the finite extension field arithmetic with polynomial and normal bases in C. We consider $m=127$, which is the specified extension degree for the 128 bit security equivalent in the RQC cryptosystem. The size of the base field is 2.

\subsection{Implementation of Polynomial Basis Operations in C} \label{s-impl-poly-fqm}
For the polynomial basis, we reference the implementation that is currently used in the RQC implementation \cite{rqc}. According to the authors it is based on the C++ library NTL.

The reference implementation uses two unsigned 64-bit integers to store a coefficient vector, one storing the lower 64 bits and the other the upper 63 bits padded with a zero bit. The authors of RQC also provide an optimized implementation using the x86 SSE instruction set, which we will not consider in this paper.

The sum of two coefficient vectors is performed with two bit-wise XOR operations, adding the lower and upper integers of the coefficient vectors, respectively.

As described in Section~\ref{s-preliminaries}, computing the multiplication of two finite extension field elements $a$ and $b$ is divided into two steps. The first step computes the unreduced product of the two binary polynomials. It is efficiently implemented by an algorithm based on the right-to-left comb method (see \cite{hankerson} Algorithm 2.36). For all 16 polynomials $u$ of degree smaller than four, the partial results $a\cdot u$ are pre-computed and stored in a lookup table with three 64bit integers per coefficient vector. This takes 28 XOR and 35 shift operations in C. The unreduced polynomial $c = a\cdot b$ is then computed using the partial results; consider \cite{hankerson, frey} for details. This takes another 29 shift operations, 24 XOR and 17 AND operations in C. The resulting coefficient vector is stored in four unsigned 64-bit integers. The second step performs the modular reduction by the irreducible polynomial $\Pi$.  As $\Pi = x^{127} + x + 1$ is a trinomial, the reduction only needs few operations.
This particular trinomial has the property 
\begin{equation*} \label{f-impl-reduce}
    x^{127+i} \equiv x^{i+1} + x^i \mod  \Pi, i=0,\dots, 125.
\end{equation*}
Additionally, the distributive property $(x^i+x^j)\mod \Pi \equiv x^i\mod \Pi+x^j\mod  \Pi$ holds. Thus, all higher order coefficients $c_i, i=127,\dots, 252$ have to be shifted to the positions $i-126$ and $i-127$, respectively, and have to be added to the lower order coefficients. As the unreduced coefficient vector is stored in 64bit segments, the computation requires slicing, concatenating and adding of segments. This is accomplished with six shift operations, six XOR operations, and one AND operation in C.

For squaring, the coefficient vector is zero-interleaved by a pre-computed look-up table that maps 8-bit integers to interleaved 16-bit integers. Hence, the input coefficient vector is sliced into 8-bit segments and the corresponding 16-bit fragments are concatenated. This takes 16 AND operations, 13 shift operations, and 6 XOR operations in C. The interleaved vector is then reduced using the same function as for the multiplication.

Elements are inverted using the extended euclidean algorithm (EEA), see \cite{hankerson}. 
As shown later, the EEA is rarely needed in the decoder and we can neglect its complexity.

\subsection{Implementation of Normal Basis Operations in C}\label{s-impl-normal-fqm}
For the extension degree $m=127$ and $q=2$ there exists no optimal normal basis, hence, we construct a low-weight normal element using the algorithm described in \cite{ash}. Our normal basis has the complexity $C_M = 501$ and is self-dual.

Similarly to the polynomial basis implementation, the vector representation $\vect{a}$ of $a\in\fqm$ can be conveniently stored in two unsigned 64-bit integers that we denote by $\vect{a}^{l}$ and $\vect{a}^{u}$, containing the lower 64 entries and the upper 63 entries of $\vect{a}$, respectively. We set the MSB of $\vect{a}^{u}$ to zero.

In this data representation the sum $\vect{c}$ of two vectors $\vect{a}$ and $\vect{b}$ is split into two XOR operations in C, that is $\vect{c}^{l} = \vect{a}^{l} \bxor \vect{b}^{l}$ and $\vect{c}^{u} = \vect{a}^{u} \bxor \vect{b}^{u}$.

The $i$th q-power of an element $a$ is given by cyclically shifting $\vect{a}$. As bits need to shift from $\vect{a}^{l}$ to $\vect{a}^{u}$ and vice versa, the shifting requires multiple operations. For $0\leq j\leq63$ we compute
\begin{equation*}
\begin{split}
	\vect{b}^l &= (\vect{a}^l \bsl j) | (\vect{a}^u \bsr 63-j) \\
	\vect{b}^u &= ((\vect{a}^u \bsl j) | (\vect{a}^l \bsr 64-j)) \band (2^{63} - 1).
\end{split}
\end{equation*}
The AND operation applies a bit mask setting the MSB of $\vect{b}^u$ to zero. We set $j=i\mod m$, as $a^{[i]} = a^{[i \mod  m]}$.
We can only shift by a maximum of 63 Bits, thus in case $j>63$ we calculate $b = a^{[j]} = a^{[j-m]}$ as
\begin{equation*}
\begin{split}
	\vect{b}^l &= (\vect{a}^l \bsr k) | (\vect{a}^u \bsl 64-k) \\
	\vect{b}^u &= ((\vect{a}^u \bsr k) | (\vect{a}^l \bsl 63-k)) \band (2^{63} - 1),
\end{split}
\end{equation*}
where $k = 127-j$.
In total, raising an element to a q-power requires four shift operations, two OR operations, and one AND operation.

The multiplication of two field elements is given in Equation  \eqref{f-norm-mul}. We use a performance optimized method developed by Ning and Yin \cite{ning}. In a first step, the shift tables $T_a$ and $T_b$ are computed for each operand, which are arrays storing $m$ elements in $\fqm$. The shift table $T_a$ contains the q-powers of $a$, i.e. the $i$th array element $T_a[i]$ of $T_a$ stores the representation vector of $a^{[-i]} = a^{[m-i]}$ for $i = 0\dots(m-1)$.
Similarly, the table $T_b$ contains the q-powers of $b$. Then, the product $\vect{c}$ is given by multiplying the shift tables \cite{ning}
\begin{equation} \label{f-impl-ning}
\vect{c} = \sum\limits_{i=0}^{m-1} \left(T_a[i] \band \sum\limits_{\mat{M}_{ij} = 1} T_b[j] \right).
\end{equation}
To save extra computational cost, we omit the masking with $2^{63}-1$ when computing the q-powers and instead set the MSB of the end result $\vect{c}$ to zero. Hence, the computation of a shift table takes $4m$ shift operations and $2m$ OR operations in C. In general, a multiplication requires the computation of two shift tables. If we multiply several times with the same operand, we can store its shift table for reuse. Equation \eqref{f-impl-ning} takes $2m$ AND operations and $2(C_M-1)$ XOR operations in C. To avoid additional overhead, we fix the indices $i$ and $j$ of the two summation operators.

For a multiplication with a q-power of the normal element $\alpha$, we provided the simplified formula $\vect{b} = \mat{M}(\vect{a}^{\bcl i})^{\bcr i}$. The multiplication table $\mat{M}$ is $C_M$-sparse, hence, it is most efficient to add only non-zero indices
\begin{equation*}
    \vect{b} = \left(\sum\limits_{j=0}^{m-1}\left(\sum\limits_{\mat{M}_{kj} = 1} \vect{a}'_k\right)2^j \right)^{\bcr i},
\end{equation*}
where $\vect{a}' = \vect{a}^{\bcl i}$.
We efficiently extract and add the indices $\vect{a}'_k$ using conditional boolean expressions in C. The entry $\vect{a}'_k$ is one if $\vect{a}'^l_k \band 2^k = 2^k, k< 64$ and $\vect{a}'^u_{k-64} \band 2^{k-64} = 2^{k-64}, k \geq 64$, respectively. The powers of two and the indices $i$ and $j$ are hard-coded as immediate operand values. The boolean assertions are combined with boolean XOR operators and the index $\vect{b}_i$ is set to one in case the if-statement's expression evaluates as true. In total, this requires two q-power operations, $C_M$ AND operations and comparisons, and $C_M - m$ boolean XOR operations in C.

For the computation of the multiplicative inverse we use an approach similar to \cite{mahmoud}. The inverse of $a$ is given by Fermat's little theorem, i.e., $a^{-1} = a^{2^m - 2}$ yielding
\begin{equation}
    a^{2^m - 2} = (a^{2^{m-1} - 1})^{[1]} = \left(\prod\limits_{i=0}^{m-2}a^{[i]}\right)^{[1]}.
\end{equation}
For $m=127$, we decompose $m-1 = 126 = 2\cdot3\cdot(1+2\cdot2\cdot(1+2\cdot2))$ and simplify the product as
\begin{equation*}
    \begin{split}
        \prod\limits_{i=0}^{m-2}a^{[i]} &= a_5 \cdot a_5^{[63]},\quad
        a_5 = a_4 \cdot a_4^{[21]} \cdot a_4^{[42]}\\
        a_4 &= a \cdot a_3 \cdot a_3^{[10]},\quad
        a_3 = a_2 \cdot a_2^{[5]}\\
        a_2 &= a^{[1]} \cdot a_1^{[2]} \cdot a_1^{[4]},\quad
        a_1 = a \cdot a^{[1]}.
    \end{split}
\end{equation*}
Thus, an inversion consists of nine multiplications and ten q-powers in $\fqm$. We store shift tables of partial results that are needed more than once. This reduces costs by nine computations of shift tables.

\section{The Welch-Berlekamp Algorithm}\label{s-wba}
The Welch-Berlekamp like Algorithm (WBA) for decoding Gabidulin codes was first presented by Pierre Loidreau in 2006 \cite{loidreau}. 
It was further improved by D. Augot et al \cite{augot}. The WBA is currently used in the RQC implementation.
In the following, we will summarize the algorithm that is optimized by Loidreau's improvement for polynomials of small degree. We will present the theoretical computational complexity and highlight differences to the RQC implementation.

\subsection{Summary of the Steps}\label{s-wba-steps}
The Welch-Berlekamp algorithm decodes by interpolating two pairs of polynomials $(P_0, Q_0)$ and $(P_1, Q_1)$.

In the initialization step two polynomials $\mathcal{A}$ and $\mathcal{I}$ are computed that evaluate to zero and interpolate $\vect{r}$ at the positions $\vect{g}_i, i=0,\dots, (k-1)$, respectively.
The two pairs are initialized as $(P_0, Q_0) = (X, 0)$ and $(P_1, Q_1) = (0, X)$.
The discrepancy vectors $\vect{u}_0$ and $\vect{u}_1$ describe the error of the interpolation. They initially evaluate as $\vect{u}_{0,i} = \mathcal{A}(\vect{g}_i)$ and $\vect{u}_{1,i} = \mathcal{I}(\vect{g}_i)-\vect{r}_i, i=0,\dots,(n-1)$, where the first $k$ entries are zero and do not need to be computed explicitly.

After initialization, the polynomials are interpolated in a for-loop with indices $l = k,\dots, (n-1)$. In every iteration, the next index $l\leq d < n$ is searched such that $\vect{u}_{1, d} \neq 0 \vee \vect{u}_{0, d} = 0$. If no such index exists, the loop is terminated early. Otherwise, the two indices $l$ and $d$ are swapped for both discrepancy vectors and the polynomials are updated. In particular, if $\vect{u}_{1, l} \neq 0$, a nominal interpolation step
\begin{equation*}
    \begin{array}{l}
P_{1}^{\prime} \longleftarrow P_{1}^2 - \frac{\vect{u}_{1, l}^{2}}{\vect{u}_{1, l}} P_{1} \\
Q_{1}^{\prime} \longleftarrow Q_{1}^2 - \frac{\vect{u}_{1, l}^{2}}{\vect{u}_{1, l}} Q_{1} \\
P_{0}^{\prime} \longleftarrow P_{0}-\frac{\vect{u}_{0, l}}{\vect{u}_{1, l}} P_{1} \\
Q_{0}^{\prime} \longleftarrow Q_{0}-\frac{\vect{u}_{0, l}}{\vect{u}_{1, l}} Q_{1}
\end{array}
\end{equation*}
and if $\vect{u}_{0, l}=0 \wedge \vect{u}_{1, l}=0$, a dummy interpolation step
\begin{equation*}
\begin{array}{l}
(P_{0}^{\prime},Q_{0}^{\prime}) \longleftarrow (P_{0},Q_{0})\\ 
(P_{1}^{\prime},Q_{1}^{\prime})  \longleftarrow (P_{1}^2, Q_{1}^2)
\end{array}
\end{equation*}
is performed. After the interpolation step, the indices of the pairs of polynomials are swapped, i.e. $(P_0, Q_0) \leftarrow (P_1', Q_1')$ and $(P_1, Q_1) \leftarrow (P_0', Q_0')$.

Next, the indices $i=l+1,\dots,(n-1)$ of the discrepancy vectors are updated. In case of nominal updates, it holds that
\begin{equation*}
    \begin{array}{l}
\vect{u}_{0,i} \longleftarrow \vect{u}_{1,i}^2+\frac{\vect{u}_{1, l}^2}{\vect{u}_{1, l}} \vect{u}_{1,i} \\
\vect{u}_{1,i} \longleftarrow \vect{u}_{0,i}+\frac{\vect{u}_{0, l}}{\vect{u}_{1, l}} \vect{u}_{1,i},
\end{array}
\end{equation*}
else, for dummy updates $\vect{u}^{\prime}_{1} = \vect{u}_{0}$ and $\vect{u}^{\prime}_{0,i} = \vect{u}^{2}_{1,i}$.

After interpolation, the decoded message $\vect{m}$ can be retrieved as the first $k$ coefficients of the polynomial $F$ which is obtained by the left Euclidean division $F = Q_{1} \backslash\left(P_{1}\cdot \mathcal{A}\right)+\mathcal{I}$.

\subsection{Implementation in RQC}\label{s-wba-implementation}
In its core, the  decoder implemented in RQC is presented by Augot et al. in \cite[Algorithm 5]{augot}. The implementation uses the optimization for polynomials of lower degree and the optimized update rule for the discrepancies as shown in \cite[Section 4.3.2]{augot}. The parameters are set to $n=113$ and $k=3$. Additionally, it has been modified to decode in constant time irrespective of the error weight. This is achieved by eliminating the early termination and dummy updates. Instead, random values are used for continuing the interpolation once the discrepancy vector $\vect{u}_1$ is all zero. The dummy interpolations are replaced by nominal interpolations.

The implementation uses the polynomial basis implementation of $\fqm$ presented in Section \ref{s-impl-poly-fqm}. It stores q-polynomials and vectors in C arrays, thus, as contiguous blocks of memory.

\subsection{Theoretical Complexity Analysis}\label{s-wba-cost}
The implementation in RQC always performs nominal updates. Thus, the upper bound of the complexity given in \cite{augot} assuming only nominal updates reflects the theoretical complexity of the constant time implementation. A summary of the complexity involved in every step is given in Table \ref{t-wba-cost}, which is based on the analysis in \cite{augot}.
\begin{table}[h]
\centering
\vspace{-1.5em}
\caption{Theoretical cost analysis of the WBA}
\label{t-wba-cost}
\vspace{-1em}
\begin{tabular}{|l|l|l|}
    \hline
               & Additions in $\mathbb{F}_{q}$ & Mult. in $\mathbb{F}_{q}$\\
    \hline
    Init. $\mathcal{A}$, $\mathcal{I}$ & $m(2k^2-2k)+2kC_{\text{inv}}$           & $(2k^2+k)m^2$ \\
                    &  $+(2k^2+k)(m^2-1)$ &  $ $\\
                    & $+(1.5k^2-0.5k)(2m-2)$  &  \\
    \hline
    Init. $\vect{u}_{0/1}$ & $m(2k-1)(n-k)$ & $(2k+1)(n-k)$ \\
    & $+(2k+1)(n-k)(m^2-1)$ & $\cdot m^2$ \\
    & $+(k-1)(n-k)(2m-2)$ &  \\
    \hline
    Up. $\vect{u}$  & $(n^2 - 2kn - n + k^2 + k)m$ &  $m^2(n^2 - 2kn $                    \\
             & $+(n^2 - 2kn - n +k^2 + k)$ & $- n +k^2 + k)$ \\
             & $\cdot (m^2-1)$ & \\
            & $+(n^2 - 2kn  + n + k^2 - k)/2 $  &   \\
            & $\cdot (2m-2)$ &  \\
    \hline
    Up. Poly. & $m(n^2 - kn)+2C_{\text{inv}}(n-k)$ & $m^2(n^2 - kn$ \\
    & $+(n^2 - kn + 2(n-k))(m^2-1) $ &  $+ 2(n-k))$\\
    & $+(k^2 - 2nk + 3k + n^2 + 2n)/2$ &  \\
    & $\cdot (2m-2)$ &  \\
    \hline
    Left div. & $(k-1)\frac{n-k}{2}m$ & $m^2(k-1)\frac{n-k}{2}$ \\
             &  $+(k-1)\frac{n-k}{2}(m^2-1)$ & \\
              & $+(n-k)(k-1)(2m-2)$ &  \\
    \hline
    Comp. $\vect{m}$ & $(k+1)m$ &  \\
    \hline
\end{tabular}
\vspace{-1em}
\end{table}

\section{The Transform Domain Decoder}\label{s-tdd}
The transform domain decoder was first presented by D. Silva and F. R. Kschischang in 2009 \cite{silva}. It is an optimization for low-rate codes, which is derived from the previously existing method based on the Berlekamp-Massey algorithm (BMA) \cite{richter}. In the following, we will summarize the steps of the algorithm, evaluate the theoretical computational complexity, and describe our implementation.
\subsection{Summary of the Steps}\label{s-tdd-steps}
The TDD decodes by determining the unique error word $\vect{e}$ of rank $\tau\leq \tau_{\text{max}}$ such that $\vect{r} = \vect{c} + \vect{e}$, where $\vect{r}$ is the received word and $\vect{c}$ is the desired code word.

The decoder has a fixed parity-check matrix with $\mat{H}_{i,j} = \alpha^{[i+j]},\, 0\leq i <  n-k, 0\leq j < m$ which only contains q-powers of $\alpha$. For decoding arbitrary Gabidulin codes defined by the partiy check matrix $\mat{H}'$, we need to apply a transformation matrix $\mat{A}\in \fq^{m\times n}$ transforming $\mat{H}' =  \mat{H} \mat{A}$. Finding $\mat{A}$ is not part of the decoding.

The TDD interprets vectors in $\fqm$ as coefficient vectors of q-polynomials. Specifically, it zero-pads $\vect{r}$, $\vect{c}$, and $\vect{e}$ and treats them as the coefficient vectors of the q-polynomials $R$, $C$ and $E$, respectively.

In the first step, we transform the received word $\vect{r}'$ by calculating $\vect{r} = \mat{A}\vect{r}'$ using the pre-computed matrix $\mat{A}$.
Then we compute the syndromes
 $\vect{s}_i = \sum_{j=0}^{m-1}\vect{r}_i\alpha^{[i+j]}$, for $i = 0,\dots,d-2$.
As $\mat{H}$ contains the q-powers of $\alpha$, the coefficients $\tilde{\vect{e}}_i$ of the q-transform of $E$ are identical with the syndromes $\vect{s}_i, i=0\dots d-2$.

Next, we determine the so-called error span polynomial $\Gamma$ of q-degree $\tau$.
We compute it with the Berlekamp-Massey algorithm solving the key-equation
\begin{equation*}
    \sum_{i=0}^{\tau} \vect{\gamma}_{i} \vect{s}_{j-i}^{[i]}=0,\quad j=\tau, \dots, d-2.
\end{equation*}
With the error span polynomial, the remaining indices $\tilde{\vect{e}}_j, j=d-1 \dots m-1$ are given explicitly by
\begin{equation*}
    \tilde{\vect{e}}_{j}=-\sum_{i=1}^{\tau} \vect{\gamma}_{i} \tilde{\vect{e}}_{j-i}^{[i]}=0, \quad j=d-1, \ldots, m-1.
\end{equation*}
The inverse q-transform yields
 $   \vect{e}_i = \sum_{j=0}^{m-1}\tilde{\vect{e}}_j\alpha^{[i+j]}$, for $i=0\dots m-1.$
Note that we choose a self-dual normal element.
The code transform is reversed yielding $\vect{e}' = \mat{A}^{\dagger}\vect{e}$ where $\mat{A}^{\dagger}$ is a left-inverse of $\mat{A}$. The retrieved code word $\vect{c}' = \vect{r}' - \vect{e}'$ is used to calculate the original message $\vect{m}'$. As every square sub-matrix of $\mat{G}'$ is invertible, we invert the sub-matrix consisting of the first $k$ columns of $\mat{G}'$. We call this inverse $\mat{G}'^{-1}_{\text{sub}}$. Similarly, we denoted the first $k$ entries of $\vect{c}'$ as $\vect{c}'_{\text{sub}}$. Then, $\vect{m}'$ is given by $\vect{c}'_{\text{sub}}\mat{G}'^{-1}_{\text{sub}}$.

\subsection{Theoretical Complexity Analysis}\label{s-tdd-cost}
To demonstrate the performance of the TDD, Silva and Kschischang evaluated the cost of the decoder by counting the number of addition and multiplications in the extension as well as the base field. In the complexity analysis they neglect the cost of that shifting operations. Table \ref{t-tdd-cost} is based on their results and summarizes the number of operations in $\fq$.
\begin{table}[h]
\centering
\vspace{-1.5em}
\caption{Theoretical cost analysis of the TDD}
\label{t-tdd-cost}
\vspace{-1em}
\begin{tabular}{|l|l|l|}
    \hline
               & Additions in $\fq$ & Multiplications in $\fq$\\
    \hline
    Code Trafo & $(n-1)m^2$           & $nm^2$                 \\
    \hline
    Syndromes  & $(n(C_M-m)+(n-1)m)$ &                      \\
               & $\cdot (d-1)$       &                      \\
    \hline
    BMA & $m(d-1)((\frac{1}{2}C_{\text{inv}}+d-2)$ & $m^2(d-1)$ \\
        & $\cdot(C_M-1)+\frac{1}{2}(d-2))$ & $\cdot(d-2+\frac{1}{2}C_{\text{inv}})$\\
    \hline
    Comp. of $\tilde{\vect{e}}$ &  $(\tau(C_M-1)+\tau-1)$ & $m^2(m-d+1)\tau$\\
            & $\cdot m(m-d+1)$  &  \\
    \hline
    Inv. q-Trafo & $nmC_M$ & \\
    \hline
    Trafo. w. $A^{\dagger}$ & $(n-1)nm$         & $n^2m$\\
    \hline
    Comp. of $\vect{m}'$ & $m(k(k-1)+(C_M-1)k^2)$ & $m^2k^2$ \\
    \hline
\end{tabular}
\vspace{-1em}
\end{table}

\subsection{Implementation in C}\label{s-tdd-implementation}
We implemented the TDD decoder for the security parameters specified in the RQC security level I, that is $n = 113$ and $k=3$. Our implementation uses the normal basis implementation presented in Section \ref{s-impl-normal-fqm}.

\section{Benchmarks and Comparison}\label{s-comp}
In this section, we compare the performance of the Welch-Berlekamp Algorithm as implemented in RQC and our implementation of the Transform Domain Decoder. We benchmark the algorithms for random Gabidulin codes, random messages, and random error words of rank $\tau_{\text{max}}$. The benchmark is executed on a 2.3 GHz Intel Core i5 processor on a single core. 
We use the gcc compiler with a -O3 optimization flag. Compiling without optimization, i.e. with flag -O0, results in a longer execution time, while the relative time differences remain approximately the same.
\subsection{Operations in the Finite Field}\label{s-comp-fqm}
We benchmark the finite field operations presented in Section \ref{s-impl-fqm} by measuring the CPU time for executing the respective function $10^6$ times.  We use the C library function \textit{clock} and calculate the difference between start and end time. Table \ref{t-comp-fqm} shows the results.
\begin{table}[h]
\centering
\vspace{-1.5em}
\caption{CPU Time [s] per $10^6$ Function Calls}
\vspace{-1em}
\begin{tabular}{|l|l|l|}
    \hline
               & Polynomial Basis & Normal Basis \\
    \hline
    add & $<5\cdot10^{-4}$  & $<5\cdot10^{-4}$ \\
    multiply     & 0.052 & 0.41 \\
    set shift table & - & 0.11 \\
    multiply shift tables & - & 0.18 \\
    multiply by $\alpha^{[i]}$& - & 0.13 \\
    q-power & - & 0.0018 \\
    square & 0.011 & 0.0018 \\
    invert & 0.53 & 2.9 \\
    \hline
\end{tabular}
\label{t-comp-fqm}
\vspace{-1em}
\end{table}
Apparently, the multiplication and inversion operation is much more efficient in the polynomial basis. Note that a multiplication in the polynomial basis theoretically requires $m^2-1 = 16128$ additions and $m^2 = 16129$ multiplications in $\fq$, thus $32257$ operations in total. A multiplication in the normal basis takes $m(C_m-1) = 63500$ additions and $m^2 = 16129$ multiplications in $\fq$ adding up to $79629$ base field operations, which is about $2.47$ times as much as the polynomial multiplication. However, the processing speed of the normal basis multiplication compared to polynomial basis multiplication is $7.9$ times lower.

The performance gap between the theoretical complexity and the run-time is even bigger for the multiplication with powers of $\alpha$. Ideally, this operation takes $C_m-m = 374$ binary additions in $\fq$ using a normal basis, while it remains a generic multiplication in the polynomial basis. Thus, the normal basis performs the operation with just $1.2\%$ of the cost of the polynomial multiplication. Nevertheless, the normal basis implementation is about $2.5$ times slower than the polynomial one. While the cost of shifting and slicing indices is neglected in the theoretical analysis, it dominates the performance for the multiplication by a q-power of $\alpha$.

\subsection{Performance of the Decoders}\label{s-decoders}
We compare the two decoders by measuring the decoding speed and the number of required operations in the finite extension field.
\subsubsection{Decoding Speed}

We estimate the total number of theoretically required additions and multiplications in $\fq$ for each of the two decoders by summing the cost of the individual steps. In total, the WBA requires about $4.18\cdot10^8$ binary additions and $4.11\cdot10^8$ binary multiplications. The TDD needs almost the equal amount of additions, that is $4.41\cdot10^8$, but only $2.20\cdot10^8$ multiplications. (Note that we neglected the cost of inversion, i.e. $C_{\text{inv}} = 0$, as there are only few inversion operations required.) Thus in theory, the TDD should be more efficient for the given parameters. In case of an optimal basis, the TDD would perform even better.

The implementation, however, shows quite a different picture. We measure the decoding time of the two decoders for $10^3$ repetitions. The WBA requires $1.75s$, while the TDD takes $7.05s$, hence, is about 4 times slower.
\subsubsection{Required Arithmetic Operations}
To provide a more detailed analysis of why the WBA performs better than the TDD, we count the number of function calls for the basic arithmetic operations in the finite extension field, see Table \ref{t-comp-decoder}.
\begin{table}[ht]
\centering
\vspace{-1.5em}
\caption{Average Number of Function Calls per Decoding}
\vspace{-1em}
\label{t-comp-decoder}
\begin{tabular}{|l|l|l|}
    \hline
               & WBA & TDD \\
    \hline
    add & 47751  & 49164 \\
    multiply     & 26021 & - \\
    set shift table & - & 12833 \\
    multiply shift tables & - & 8699 \\
    multiply by $\alpha^{[i]}$& - & 28321 \\
    q-power & - & 3960 \\
    square & 13547 & 3080 \\
    invert & 114 & 55 \\
    \hline
\end{tabular}
\vspace{-1em}
\end{table}
The TDD requires less generic multiplications, less squares, and less inversions than the WBA. However, the TDD multiplies by q-powers of $\alpha$ a lot. As we have seen above, this operation in particular performs worse than estimated in theory.

\section{Conclusion and future work}\label{s-conclusion}
In this paper, we have compared two implementations of finite extension fields, one based on a polynomial basis, the other on a normal basis representation. Our benchmarks have shown that the software implementation of the polynomial basis outperforms the normal basis. The theoretical assumptions that shift operations and vector indexing operations are negligible do not hold for the software implementation. In particular, the multiplication with $q$-powers of $\alpha$ performs much worse than expected. Our benchmarks can be translated to estimate the time complexity of other algorithms using finite extension fields.

Based on the two finite extension field implementations, we have compared the Welch-Berlekamp algorithm and the Transform Domain Decoder. Considering the theoretical complexity, the TDD outperforms the WBA for the given parameter set. The benchmarks of our implementation in C, however, show that the theoretical assumptions are not good enough for predicting the performance of the implementation. In fact, the WBA is four times more time efficient than the TDD.

The parameter set given in RQC prevented the usage of an optimal normal basis. Hence, future research could involve re-implementing the finite extension field for choices of $m$, where an optimal self-dual normal basis exists.

\bibliographystyle{ieeetr}

\end{document}